\documentclass[%
preprint,
 amsmath,amssymb,
prb,
]{revtex4-1}

\usepackage{graphicx}
\usepackage{dcolumn}
\usepackage{bm}

\begin{document}

\preprint{APS/123-QED}

\title{One step model of photo-emission at finite temperatures: spin fluctuations of Fe(001)}
\author{J\'an Min\'ar}
\affiliation{New Technologies-Research Center, University of West Bohemia,
  Univerzitn\'i 8, 306 14 Plze\v{n}, Czech Republic.}
  \email{jminar@ntc.zcu.cz}
\author{Sergey Mankovsky}
\affiliation{Department of Chemistry,
LMU Munich,  Butenandtstra\ss{}e 11, 81377 M\"unchen, Germany.}
\author{J\"urgen Braun}
\affiliation{Department of Chemistry,
LMU Munich,  Butenandtstra\ss{}e 11, 81377 M\"unchen, Germany.}
\author{Hubert Ebert}
\affiliation{Department of Chemistry,
LMU Munich,  Butenandtstra\ss{}e 11, 81377 M\"unchen, Germany.}

\date{\today}

\begin{abstract}
Various technical developments extended the potential of angle-resolved photoemission (ARPES) tremendously during the last
twenty years. In particular improved momentum, energy and spin resolution as well as the use of photon energies from few eV
up to several keV makes ARPES a rather unique tool to investigate the electronic properties of solids and surfaces.
With our work we present a generalization of the state of the art description of the photoemission process, the so called
one-step model that describes excitation, transport to the surface and escape into the vacuum in a coherent way. In particular,
we present a theoretical description of temperature-dependent ARPES with a special emphasis on spin fluctuations.
Finite temperature effects are included within the so called alloy analogy model which is based on the coherent potential
approximation and this way allows to describe uncorrelated lattice vibrations in combination with spin fluctuations quantitatively
on the same level of accuracy. To demonstrate the applicability of our approach a corresponding numerical analysis has been
applied to spin- and angle-resolved photoemission of Fe(100) at finite
temperatures.
\end{abstract}

\maketitle

\section{Introduction}
The experimental and theoretical studies on itinerant electron ferromagnetism address one of the crucial problems in condensed matter
physics (for reviews, see  Ref.\ \onlinecite{vollhardt1999metallic}). One of the most important  experimental tools to get direct insight
into the electronic structure of solids and surfaces is angle-resolved
photoemission spectroscopy (ARPES). In particular, spin- and angle-resolved
photoemission (SARPES) has been developed into a powerful method to study surface and thin film magnetism \cite{johnson1997spin}. Very
recently this technique has been used extensively to investigate the
topological properties of solid state materials \cite{dil2019spin}.
In the 1980-ies first experimental studies using SARPES had been devoted to probe the existence of local magnetic moments at temperatures
close and above the Curie temperature $T_{\rm C}$. Pioneering SARPES experiments had been performed in
particular by Kisker et al. \cite{kisker1984photoemission,kisker1985spin,kisker1984temperature} on Fe(001).
At that time two contrary models had been proposed to describe the
ferromagnetic to paramagnetic transition at the critical temperature.
On the one-hand side,  the so-called
Stoner model proposed the breakdown of the exchange splitting of bands
 leading this way to the non-magnetic phase. On the other hand the existence
of local fluctuating magnetic moments above the Curie temperature
according to the Heisenberg model was suggested. SARPES studies on magnetic
transition metals (Fe and Co) were able to clearly
identify the exchange-split bands at lower temperatures and
fluctuating moments at high temperatures \cite{johnson1997spin}.

Unfortunately, after the pioneering SARPES experiments at elevated temperatures, most of the more recent SARPES studies were done at room
or even at very low temperatures for a variety of materials including superconductors, topological materials etc. \cite{DHS03,dil2019spin,BME18}.
The main reason for this is found in a possible contamination of the electron analyzer and UHV chamber after heating of the thin
film samples which leads to a significant decrease of the pressure in the UHV chamber. However, the thermal vibrations in combination with spin
fluctuations turned out to be a very important issue for photoemission spectra measured at high photon energy ranging from soft to hard-X-rays
\cite{woicik2016hard,fadley2012looking,GPU+11,VMB+08, PMS+08}. Going to higher photon energies has the advantage of a longer inelastic mean free path of the
photoelectons and turns ARPES to a bulk sensitive technique. However, higher photon energies  challenge the interpretation of the corresponding
experimental spectra. In particular, even at very low temperatures (tenths of a Kelvin), indirect
transitions occur which in
consequence lead to the  XPS limit. The corresponding averaging over the Brillouin zone leads to density of states like spectra
for any emission angle and the access to the ground state band structure is lost \cite{MBE13a}. Finally, spin fluctuations play an important
role in the description of ultrafast processes measured by pump-probe angle-resolved photoemission and two photon photoemission spectroscopy.
Absorption of a very intense pump-pulse leads in the first femtoseconds to the increase of the electronic temperature and after several
hundreds of femtoseconds the energy is dissipated into the lattice. Very recently, first time-dependent SARPES measurements have been performed
for topological insulators \cite{CCB+15}. Furthermore, Eich et al. performed a detailed study on possible ultrafast demagnetization processes in
ferromagnetic transition metals by SARPES \cite{EPR+17}.

It is well known that density functional theory (DFT) in its local spin-density approximation is able to describe quantitatively the ground
state and magnetic properties of transition metals at $T=0$~K. This rigorous description can be extended also to finite temperatures. The most
common multi-scale
approach in this direction
is based on the calculation of the so called exchange coupling constants \cite{LKAG87} for a classical Heisenberg
model on the basis of DFT
and to perform subesquent Monte Carlo or spin dynamics simulations.
On the other hand, it has been realized since many years that locally fluctuating magnetic moments are a consequence of
local electronic correlations. A very successful method to go beyond the DFT-LSDA scheme is the dynamical mean field theory (DMFT) in combination
with DFT. Liechtenstein et al. showed that such a DFT+DMFT approach can quantitatively describe temperature-dependent magnetism in Fe and Ni
\cite{LKK01}.
However such an approach does not take into account lattice vibrations which are present at all finite temperatures.
On the other hand,
a scheme to deal with thermal lattice vibrations
is provided by the so-called
alloy analogy model \cite{EMKK11} that
takes the necessary thermal average by means
of the coherent potential approximation (CPA)
alloy theory. This approach was already
applied successfully to deal with
ARPES of non-magnetic materials
at finite temperatures \cite{BMM+13}.
In addition,
following the orignal idea behind the
 alloy analogy model  it was
extended to  account for thermally induced spin fluctuations in magnetic materials \cite{EMC+15}
as well. This opens the combination with various models
to deal with thermal spin fluctuations
as for example the  disordered local moment
approach \cite{SGP+84,SBS+14}.
Another advantage of the approach is its
possible combination with methods describing local correlations as for
example LSDA+U and LSDA+DMFT. This was demonstrated recently for Gd,
where temperature dependence of the longitudinal resistivity and the
anomalous Hall effect was studied \cite{CKM17}.

It is widely accepted to interpret a measured photoelectron spectrum by referring to the results of band-structure calculations. Such an
interpretation is questionable for moderately and even more for strongly correlated systems.
On the other hand,
the most reliable theoretical approach to interprete
ARPES spectsa is provided by the so-called one-step model of photoemission. This approach was formulated first by Pendry and co-workers
\cite{Pen76,HPT80} in the framework of multiple scattering theory and has been recently generalized to include various aspects like e.g. disorder,
lattice vibrations, electronic correlations, the fully relativistic spin-density matrix formulation and time-dependent pump-probe aspects
\cite{BME18,MBME11,MBE13}. However this scheme did not allow up to now to consider temperature-dependent spin fluctuations in combination with lattice
vibrations. In this paper we generalize the one-step model of photoemission in order to include spin-fluctuations and lattice vibrations on
the same level of accuracy within the framework of the alloy analogy model.

The paper is organized as follows: In Sec.~\ref{sec:theory} we describe the theoretical approach, the so called alloy analogy model, which has
been applied to the one-step model of photoemission in the framework of the SPR-KKR method. In Secs.~\ref{sec:dos} and \ref{sec:fe} we apply
this formalism and  calculate temperature-dependent, spin-polarized ARPES spectra for Fe(001). In Sec.~\ref{sec:sum} we summarize our results.

\section{Theoretical approach: thermal effects}
\label{sec:theory}
Considering the electronic structure of a magnetic solid at finite temperature, its modification due to thermal lattice and magnetic excitations
has to be taken into account. The present approach is based on the adiabatic treatment of the non-correlated localized thermal
displacements of atoms from their equilibrium positions (thermal lattice vibrations) in combination with a tilt of the local magnetic
moments away from their orientation in the ground state
 (thermal spin fluctuations). Multiple scattering theory allows to describe uncorrelated
local thermal vibrations and spin fluctuations within the single-site CPA alloy theory. This implies the reduction of the calculation
of a thermal average to the calculation of the configurational average in complete analogy to
the averaging  for random, substitutional
 alloy systems. The impact of thermal
effects on the electronic structure, taken into account within such an approach, was discussed previously in order to describe the
temperature dependent transport properties and Gilbert damping in magnetic systems \cite{MBE13}. The impact of the thermal lattice
vibrations was also studied in calculations of temperature-dependent photoemission of non-magnetic systems \cite{BMM+13}, however
the inclusion of the thermal spin fluctuations for ferromagnetic systems is missing and in the following we generalize the one-step
model of photoemission accordingly.

\subsection{Alloy analogy model}
\label{sec:aam}

Within the alloy analogy model, lattice vibrations are described by a discrete set of $N_{v}$ displacement vectors $\Delta \vec{R}^q_v(T)$ for each atom in
the unit cell. The temperature dependent amplitude of the displacements is taken to be equal to the root mean  square displacement
$(\langle u^2\rangle_T)^{1/2}$, $|\Delta \vec{R}^q _v(T)| = \langle u_q ^2 \rangle _T ^{1/2}$, with the probabilities $x_v = 1/N_{v}$
($v=1,..,N_{v}$). $[\langle u_q^2\rangle _T]^{1/2}$ is evaluated here within the Debye model with the Debye temperature $\Theta_D$ taken
from  experiment.

Using the rigid muffin-tin approximation \cite{PZDS97,Lod76}, the displaced atomic potential is associated with a corresponding
single-site t-matrix $ \underline{t}$
that has to be referred
with respect to the common global frame of reference. This quantity is obtained by  a coordinate
 transformation from local
single-site t-matrix $ \underline{t}^{\rm loc}$
via the expression:
%
\begin{equation}
\label{eq:U-trans}
\underline{t} = \underline{U}(\Delta \vec{R})\,\underline{t}^{\rm loc}\,
                     \underline{U}(\Delta \vec{R})^{-1} \;.
\end{equation}
%
In the following the underline symbol represents a matrix in the angular momentum representation. In
the fully relativistic formulation case, as
adopted  here, this implies a labelling
of the matrix elements with the relativistic
quantum numbers $\Lambda=\kappa,\mu$ \cite{Ros61}. The so-called U-transformation
matrix $\underline{U}(\vec{s})$
in Eq.\ (\ref{eq:U-trans})
is given in its non-relativistic form by:\cite{Lod76,PZDS97}
%
\begin{equation}
\label{eq:U-trans-matrix}
U_{LL'}(\vec{s}) =
4\pi \sum_{L''}i^{l+l''-l'}\, C_{LL'L''}\, j_{l''}(|\vec{s}|k)\, Y_{L''}(\hat{s})
\;.
\end{equation}
%
Here $L=(l,m)$ represents the non-relativistic angular momentum quantum numbers, $j_{l}(x)$ is a spherical Bessel function,
$Y_{L}(\hat{r})$ a real spherical harmonics, $C_{LL'L''}$ a corresponding Gaunt number and $k=\sqrt{E}$ is the electronic wave
vector. The relativistic version of the U-matrix is obtained by a standard Clebsch-Gordan transformation.\cite{Ros61}

\medskip

To account for the impact of disorder caused by thermal spin fluctuations, the continuous distribution $P(\hat{e})$
for the orientation of local magnetic moments is replaced by a discrete set of orientation vectors $\hat{e}_f$ (with $f=1,...,N_f$)
occurring with a probability $x_f$. The configurational average for this discrete set of orientations is made using the CPA leading
to a periodic effective medium.

The rigid spin approximation \cite{LKAG87} used in the calculations implies that the spin-dependent part $B_{\rm xc}$ of the
exchange-correlation potential does not change for the local frame of reference fixed to the magnetic moment when the moment
is oriented along an orientation vector $\hat{e}_f$. As a result, the single-site t-matrix  $ \underline{t}_f^{\rm loc}$  considered in the
local frame is the same  for all orientation vectors. With respect to the common global frame that is used to deal with the
multiple scattering problem (see Eq.~(\ref{eq:CPA3})) the t-matrix for a given orientation vector $(\hat{e}_f)$ is determined by:
%
\begin{equation}
\label{eq:R-trans}
\underline{t}_f = \underline{R}(\hat{e}_f)\,\underline{t}^{\rm loc}\,
                \underline{R}(\hat{e}_f)^{-1} \;,
\end{equation}
%
with the transformation from the local to the global frame of reference expressed by the rotation matrices $ \underline{R}_f =
\underline{R}(\hat{e}_f)$.\cite{Ros61} The temperature dependent probability $x_f=x(\hat{e}_f)$ for each orientation $\hat{e}_f$ and
an appropriate Weiss field parameter $w(T)$
is given by the expression \cite{Tik64}:
\begin{equation}
\label{eq:xf}
       x_f = \frac{\exp(-w(T) \hat{e}_{z} \cdot \hat{e}_{f}/kT)}
  {\sum_{f'} \exp(-w(T) \hat{e}_{z} \cdot \hat{e}_{f'}/kT)}   \;.
\end{equation}

The various types of disorder discussed above may be combined with each other as well as with chemical i.e.\  substitutional
disorder. In the most general case a pseudo-component $(vft)$ is characterized by its chemical  atomic type $t$, the spin
fluctuation $f$ and lattice displacement $v$. Using the rigid muffin-tin and rigid spin approximations, the single-site
t-matrix $t^{\rm loc}_t $ in the local frame is independent from the orientation vector $\hat{e}_f$ and displacement vector
$\Delta \vec{R}_{v}$, and coincides with $ \underline{t}_t$ for the atomic type $t$. With respect to the common global frame
one has accordingly the t-matrix:
%
\begin{equation}
\label{eq:tvft}
 \underline{t}_{vft} =     \underline{U}(\Delta \vec{R}_{v})\,
                     \underline{R}(\hat{e}_{f})\,
                     \underline{t}_t \,
                     \underline{R}(\hat{e}_{f})^{-1}
                     \underline{U}(\Delta \vec{R}_{v})^{-1}
\;.
\end{equation}
%
With this the resulting CPA equations are identical to the standard CPA Eqs.~(\ref{eq:CPA1}) to (\ref{eq:CPA3}) below with the index $t$  identifying atom types replaced by the
combined index $(vft)$. The corresponding pseudo-concentration $x_{vft}$ combines the concentration $x_t$ of the atomic type $t$ with
the probability for the orientation vector $\hat{e}_f$ and displacement vector $\Delta \vec{R}_{v}$. This leads to the site diagonal
configurational average which can be determined by solving the
multi-component CPA equations \cite{FS80}:
%
\begin{eqnarray}
\label{eq:CPA1}
\underline{\tau}_{{\rm CPA}} &= &
\sum_{t}
x_{t} \underline{\tau}_{vft}
\\
%
\label{eq:CPA2}
\underline{\tau}_{t}& = &
\big[
    (\underline{t}_{vft})^{-1}
-   (\underline{t}_{{\rm CPA}})^{-1}
+   (\underline{\tau}_{{\rm CPA}})^{-1}
\big]^{-1}
\\
\label{eq:CPA3}
\underline{\tau}_{{\rm CPA}}
 & = & \frac{1}{\Omega_{{\rm BZ}}}
 \int_{\Omega_{\rm BZ} } d^{3}k
\left[ (\underline{t}_{{\rm CPA}})^{-1}
      - \underline{G}(\vec{k},E)  \right]^{-1}  \; ,
\end{eqnarray}
%
where again the underline symbol indicates matrices with respect to the combined index $ \Lambda$.

\subsection{One step model of ARPES}

The main idea of the one-step model of photoemission is to describe the excitation process, the transport of the photoelectron
to the surface as well as the escape into the vacuum
in a coherent way as a single quantum mechanical process \cite{Pen76}. The one-step model of ARPES
is based on the Fermi's golden rule and  was originally implemented for ordered surfaces using the multiple scattering KKR formalism
(for more details see
review in Ref.\ \onlinecite{Bra96}). This approach has been generalized to describe photoemission of disordered alloys by means
of the CPA \cite{DTLN83,BMM+10}. Recently it was extended to deal with thermal lattice vibration effects exploiting the  alloy
analogy model  described above. This approach was successfully applied to describe indirect transitions which occur in soft- and hard-X-ray
photoemission \cite{BMM+13}. Based on the CPA approach the temperature-dependent spin-density matrix $\rho$
for a given kinetic energy
$\epsilon_f$ and wave vector ${\bf k}_{\|}$ can be written in the following form:
\begin{eqnarray}
\langle{\overline\rho}_{ss'} (\epsilon_f,{\bf k}_{\|},T)\rangle~\propto~&&
\langle{\overline\rho}^{at}_{ss'}(\epsilon_f,{\bf k}_{\|},T)\rangle
+\langle{\overline\rho}^{c}_{ss'} (\epsilon_f, {\bf k}_{\|},T)\rangle \nonumber \\
&+&\langle{\overline\rho}^{inc}_{ss'} (\epsilon_f,{\bf k}_{\|},T)\rangle
+\langle{\overline\rho}^{surf}_{ss'}(\epsilon_f,{\bf k}_{\|},T)\rangle,
\end{eqnarray}
with  a purely atomic part ($at$), a coherent part ($c$) with multiple scattering involved and an incoherent ($inc$) part as described in detail
in Refs.\ \onlinecite{GDG89} and
\onlinecite{Dur81} in the context of chemical disorder in alloys. The third, incoherent contribution which appears due to the CPA-averaging procedure
represents an on-site quantity that behaves DOS-like \cite{GDG89}. The last contribution is the surface ($surf$) part of the spin-density matrix.
As dispersing and non-dispersing contributions are clearly distinguishable we can define the spin-density matrix which describes the angle-integrated
(AIPES) photoemission
\begin{eqnarray}
\langle{\overline\rho}^{\rm AIPES}_{ss'} (\epsilon_f,{\bf k}_{\|},T)\rangle~\sim~
\langle{\overline\rho}^{at}_{ss'}(\epsilon_f,{\bf k}_{\|},T)\rangle+\langle{\overline\rho}^{inc}_{ss'} (\epsilon_f, {\bf k}_{\|},T)\rangle+
\langle{\overline\rho}^{surf}_{ss'}(\epsilon_f,{\bf k}_{\|},T)\rangle~,
\end{eqnarray}
where the ${\bf k}$-dependence in the atomic and incoherent contributions is only due to the final state. A ${\bf k}$-averaging is not necessary
because the ${\bf k}$-dependence of the (SP)LEED-type final state is very weak and can be neglected in explicit calculations. Furthermore,
when using  the single-scatterer approximation for the final state the ${\bf k}$-dependence completely vanishes. This way a direct comparison to
corresponding measurements is given in both cases.

In terms of the spin-density matrix $\rho$
the intensity of the photocurrent
can be written as:
\begin{equation}
I(\epsilon_f,{\bf k}_{\|},T)~=~Tr \left(~\rho_{ss'}(\epsilon_f,{\bf k}_{\|},T)~\right)~,
\label{eq:spind3}
\end{equation}
with the corresponding spin polarization vector given by:
\begin{equation}
{\bf P}~=~\frac{1}{I}~Tr~\left(~ \mbox{\boldmath $\sigma$}~\rho~\right)~.
\label{eq:spind4}
\end{equation}
Finally, the spin-projected photocurrent is obtained from the following expression:
\begin{equation}
I^{\pm}_{{\bf n}}~=~\frac{1}{2}~\left(~1~\pm~{\bf n} \cdot {\bf P}~\right)~,
\label{eq:spind5}
\end{equation}
with the spin polarization $(\pm)$ referring to the vector $\bf n$.

Within our approach, we aim on a generalized spin-density matrix formalism for the photo current to  include spin fluctuations and
thermal vibrations on the same level of accuracy. The formalism presented in section \ref{sec:aam} provides us with the
temperature-dependent single-site scattering matrix $\underline{t}_{vft}$ which enters the  multiple scattering KKR formalism to
calculate the photocurrent $I(\epsilon_f,{\bf k}_{\|},T)$. (A detailed description of the generalized fully relativistic one-step model for
disordered magnetic alloys can be found in
Ref.\  \onlinecite{BME18}). Special care has to be taken concerning the temperature-dependent averaging
of the photoemission matrix elements, in contrast to the previous work which did not account for spin fluctuations \cite{BMM+13}.
Within the above mentioned rigid spin approximation \cite{LKG84}, the regular $\underline M^{\rm loc}_{i'}$ and irregular
$\underline I^{\rm loc}_{i',j'}$ dipole matrix transition elements are first calculated for the local frame of reference fixed to
the magnetic moment when the moment is oriented along an orientation vector $\hat{e}_f$ with the components
$i'$ and $j'$ of the  light
polarization vector
referred to the  local frame of reference
($x',y',z'$) with  $\hat{e}_{z'}= \hat{e}_f$ . In the case of spin fluctuations, the transformation
of the matrix elements into the global frame of reference includes also a rotation of the polarization. For the regular matrix elements
one finds:
\begin{equation}
  \underline{M}^{\rm vft}_{i}=\sum_{i'}D_{i i'}(\hat{e}_{f})\,
  \underline{U}(\Delta \vec{R}_{v})\,
  \underline{R}(\hat{e}_{f})\,
  \underline{M}^{\rm loc}_{i'} \,
     \underline{R}(\hat{e}_{f})^{-1}
     \underline{U}(\Delta \vec{R}_{v})^{-1}
\;,
\end{equation}
%
and for the irregular matrix elements
one has accordingly:
\begin{equation}
  \underline{I}^{\rm vft}_{ij}=\sum_{i' j'}D_{i i'}(\hat{e}_{f})\,
  D_{j j'}(\hat{e}_{f})\,
  \underline{U}(\Delta \vec{R}_{v})\,
  \underline{R}(\hat{e}_{f})\,
  \underline{I}^{\rm loc}_{i' j'} \,
     \underline{R}(\hat{e}_{f})^{-1}
     \underline{U}(\Delta \vec{R}_{v})^{-1}
\;,
\end{equation}
where the $3\times3$ matrix $D_{ij}$ represents the transformation of the polarization vector of the light from the local to the global frame of reference.

\section{Computational details}

The electronic structure of the investigated ferromagnet
bcc Fe, has been calculated self-consistently
using the SPR-KKR band structure method \cite{SPR-KKR7.7,EKM11}. For the
LSDA exchange-correlation potential the parametrization as given
by Vosko et al. \cite{VWN80} has been used and  the experimental lattice
parameter have been taken.
For the angular momentum
expansion within the KKR multiple-scattering
method a cutoff of $l_{max}= 3$ was used.
 The temperature effects are
treated within the alloy analogy scheme based on the CPA
alloy theory. For the description of the magnetic spin fluctuations
the temperature-dependent magnetization data were
taken from experimental magnetization curves \cite{CG71}
 and the lattice displacements as a function of temperature has been calculated using the Debye temperature of $T=420$~K.
In addition to the LSDA calculations, a charge
and self-energy self-consistent LSDA+DMFT scheme for
correlated systems based on the KKR approach \cite{Min11,MCP+05}
has been used. The many-body effects are described by means
of dynamical mean field theory (DMFT) \cite{Hel07} and the
relativistic version of the so-called spin-polarized T-matrix
fluctuation exchange approximation \cite{PKL05,MMC+09} impurity solver
was used. The realistic multiorbital interaction has been parametrized
by the average screened Coulomb interaction $U$
and the Hund exchange interaction $J$. In our calculations of bcc Fe we used values for the Coulomb
parameter $U=1.5$~eV and $J=0.9$~eV as found by our previous ARPES studies on Fe \cite{SFB+09,SBM+12}.

In a second step the self consistent potential and DMFT self energy  for bcc Fe has been used to calculate the photoemission response from the Fe(001) surface by means of the one step model of photoemission as presented above.

\section{Results and Discussion}
\subsection{Temperature dependent ground state}
\label{sec:dos}

First, let's discuss the impact of thermal lattice vibration and spin
fluctuations on the ground state electronic structure of a magnetic solid,
focusing on the temperature induced modification of the density of states
(DOS). In an  ordered material, the spin (s) resolved density of
states is represented by the sum $n_s(E) = \frac{1}{N}
\sum_{\vec{k}} \delta(E - E_s(\vec{k}))$, with $E_s(\vec{k})$ the
energies of the electron states characterized by an infinite life time in
the case of $T = 0$~K. On the other hand, at  a finite temperature, $T > 0$
K, the electron scattering due to thermally induced lattice vibrations and spin
fluctuations leads to a finite life time of the electronic states which
can be accounted for within the KKR Green function formalism by giving the total
DOS in terms of the Green function as follows
\begin{eqnarray}
\label{eq:DOS}
n(E) &= & -\frac{1}{\pi} \mbox{Im } \mbox{Trace } \underline{G}(E) \;.
\end{eqnarray}
Thermally induced lattice vibrations are treated here as random atomic
displacements from the equilibrium positions, with the amplitude
dependent on temperature. The same holds for  the temperature induced
tilting of the atomic spin moments. This creates a thermal disorder in
the atomic positions and spin orientations having a similar
impact on the electronic structure as chemical disorder in an alloy. In particular, it causes a broadening of the electronic states and a
change of the exchange splitting of the states with opposite spin
direction. Using the alloy analogy formalism described above,
the Green function of the system, represented within multiple scattering theory
 is given in terms of the configurational average of the
scattering path operator $\underline{\tau}_{CPA}$ given by Eqs.\ (\ref{eq:CPA1}) to  (\ref{eq:CPA3}).

As it will be shown below, spin fluctuations have a dominating
contribution to the thermally induced modification of electronic structure  when the
temperature approaches the
critical temperature  $T_{\rm C}$, where a transition to
the paramagnetic (PM) state  occurs. Thus, focusing on thermal spin
fluctuations only, the scattering path operator averaged over spin
fluctuations at a given temperature can be written as follows
$\underline{\tau}_{CPA} =  \sum_{f} x_{f} \underline{\tau}_{f}$, where
$\underline{\tau}_{f}$ is associated with the spin orientation
$\hat{e}_f$, giving access to a corresponding DOS contribution $n_{f,s}(E)$.
The DOS $n^{loc}_{f,s}(E)$  projected on spin $s$ evaluated in the local frame
of reference with $\hat{e}_{z'} = \hat{e}_f$ is different for different
spin channels in the case of a non-zero local magnetic moment. This holds even for the PM (i.e.\ magnetically disordered) state  with
$\langle \hat{m} \rangle = 0$
in case of a non-vanishing local moment
above $T_{\rm C}$ as it occurs, e.g.\ for bcc Fe.
However, the average
spin-projected DOS functions calculated
 for
the PM state in a common global frame of reference are equal; i.e. one has  $\langle n_{\uparrow}\rangle (E) = \langle
n_{\downarrow}\rangle (E)$.
Here, the  indices ${\uparrow}$ and ${\downarrow}$
stand for a
spin orientation along the global $\hat{e}_{z}$ direction and
opposite to it, respectively. Due to random orientation of the atomic
spin magnetic moments in the system, the $n_{+}$ and $n_{-}$ DOS
projections are contributed equally by the electronic states
characterized by different spin quantum numbers, implying mixed-spin
character of the electronic states in such a system.

\begin{figure}[h]
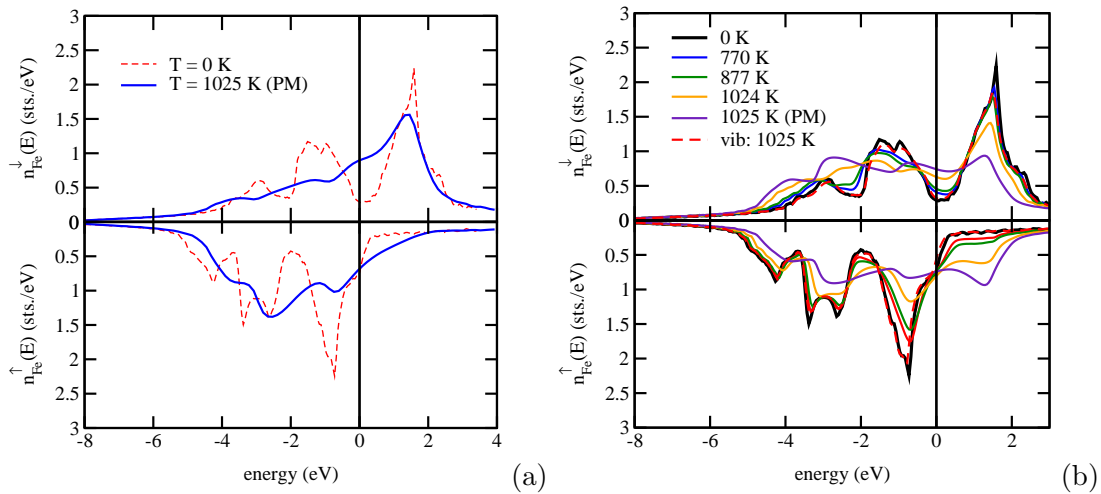

 \includegraphics[width=0.4\textwidth,angle=0,clip]{Fe_DOS_T_local.eps}\;(a)
 \includegraphics[width=0.4\textwidth,angle=0,clip]{Fe_DOS_T_global_mod.eps}\;(b)
 \caption{\label{fig:DOS_T} Total spin resolved DOS for bcc Fe in the local (a)
   and the global (b) frames of reference.
  }
 \end{figure}

Fig. \ref{fig:DOS_T} (a) represents the DOS for bcc Fe calculated for
the PM state ($\langle \hat{m} \rangle = 0$) in the
local frame of reference (solid line), averaged over all possible
orientations of the magnetic moment. This result is compared with the
DOS at $T = 0 K$. One can see first of all a  finite exchange
splitting of the majority and minority spin states at $T > T_C$. The
main temperature effect is a significant broadening of the energy bands
when compared to the case of $T = 0 K$.
However, in the global frame of reference the difference between the
majority and minority-spin states decreases approaching the
critical temperature $T_C = 1024 K$. Above $T_{\rm C}$, in the PM
state, the difference vanishes between the DOS for different spin
channels. However, this is not the case when only thermal lattice
vibrations are taken into account (dashed line in Fig. \ref{fig:DOS_T}
(a) for $T = 1025$~K). In this case only a weak broadening of the energy
bands occurs, which is much weaker when compared to that due to spin
fluctuations.

 \subsection{Angle resolved photoemission of bcc Fe(001)}
 \label{sec:fe}
Although a large number of experimental spin-resolved ARPES studies on ferromagnetic transition metals are present in the
literature, corresponding data for high temperatures are very rare. Experimental temperature-dependent studies have been
carried out predominantly for Fe and Ni in the middle of 1980-ies (for review see Ref.\ \onlinecite{johnson1997spin}). On the other hand, there have been
several attempts to account for temperature-dependent ARPES within various different theoretical frameworks such as dynamical
mean field theory \cite{LKK01}, or the disorder local moment approach \cite{DSG84}. However,
most theoretical models were limited either to $T=0$~K or to temperatures above the critical temperature $T_{\rm C}$, and are based on the ground state electronic structure only.
This way these approaches are ignoring matrix element, surface and final state effects. Therefore the question whether ARPES can distinguish between
the different models describing finite temperature spin correlations, as the Stoner or Heisenberg model, is still open
\cite{EPR+17}. The alloy analogy model in combination with the one-step model of photoemission, presented in Sec.~\ref{sec:theory},
allows to describe all the mentioned effects on the same level of accuracy. As a first illustration of an application
 of this approach
we discuss results for temperature-dependent spin-resolved ARPES on Fe(001) and compare the calculated spactra with corresponding
experimental data stemming from Kisker et al. \cite{kisker1985spin}.
 \begin{figure}[h]
  \includegraphics[width=1.0\textwidth,angle=0,clip]{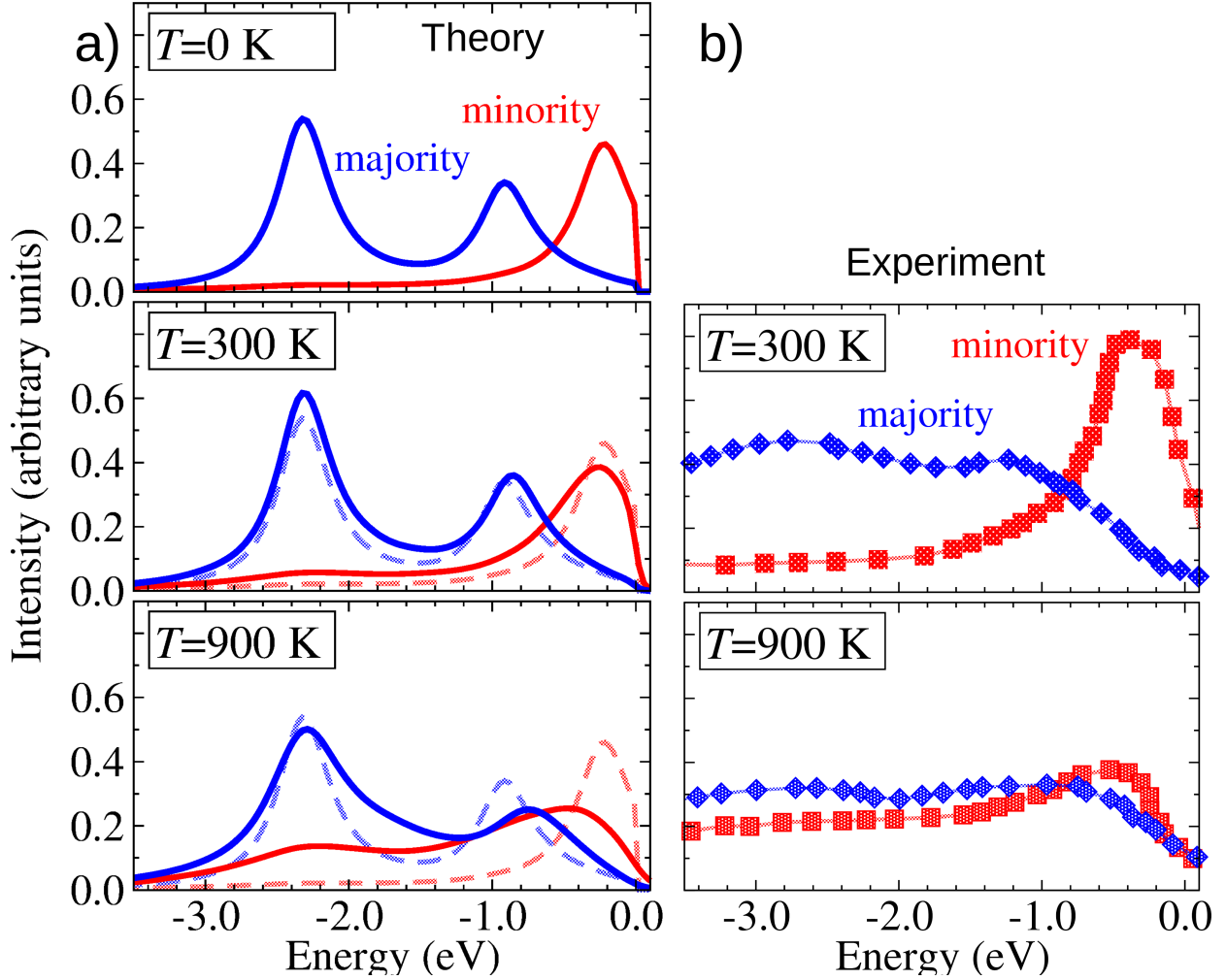}
 \caption{\label{fig:NormalEm_LDA} Comparison between experimental (right panel) and  theoretical LSDA based spectra (left panel) for temperature dependent
   spin resolved photoemission with at $E_{\rm phot}=60$~eV and normal
   emission.  The dashed lines are spectra calculated for $T=0$~K. }
 \end{figure}


In Fig.~\ref{fig:NormalEm_LDA} we compare experimental and theoretical LSDA based spin-resolved photoemission data for three
different temperatures, namely $T=0$, $300$ and $900$~K respectively. The data for $0$~K  are seen as a reference obtained by using the standard
one-step model of photoemission scheme. All spectra have been calculated for normal emission geometry assuming s-polarized light with $60$~eV
photon energy.  Prior to these calculations we performed a photon energy scan ($k_z$-scan) in order to identify the $k_z$ position in
the Brillouin zone. Due to the LSDA approximation the final states are
usually shifted somewhat in energy
with respect to the experimental spectra. In the
case of Fe the photon energy of $60$~eV corresponds to emission from the $\Gamma$ point. The spin-resolved spectra reveal three main transitions
with  bulk states as initial states: a minority peak close to the Fermi level and a majority peak at $-2.4$~eV binding energy having both T$_{2g}$ symmetry.
The majority peak at $-0.9$~eV binding energy has E$_g$ symmetry. This transition should  be suppressed by using s-polarized light due to the selection rules. However,
as mentioned by Kisker et al. \cite{kisker1985spin} due to the finite
acceptance angle of the analyzer this transition has nevertheless been observed
in the corresponding measurements. In addition a majority peak around $-0.9$~eV shows up with strong surface character and in fact
it is a mixture of an E$_g$-like state and a surface d-like resonance. The minority surface states of Fe(001) close to the Fermi
have been studied in  detail in the past \cite{plucinski2009surface} but could not be
resolved in Kisker's work
 due to the limited experimental resolution.

In Fig.~\ref{fig:NormalEm_LDA} (lower panel) results of finite temperature calculations (see Sec.~\ref{sec:theory}) are compared
with corresponding experimental data.
As a refernce,  calculated spectra for $T=0$~K
are given by dashed lines.
 Obviously, we obtained reasonable
agreement with the experimental spectra. At $T=900$~K the magnetization of Fe is decreased to roughly about 60\% of the value at $T=300$~K.
As one can see, at high temperature the E$_{g}$ states are shifted towards the Fermi level.
The exchange splitting of the  T$_{2g}$ states
 is reduced but still it
remains considerably high.
In particular, not only a reduction of the exchange splitting
is observed but also an increase of the minority peak intensity at
$-2.5$ and $-0.9$~eV
is found in accordance with the experimental
findings. This  results from an
 increasing contribution from the
majority spin states in line with the discussion in Sec.~\ref{sec:dos}.
The overall reduction in the minority spin
 intensities at
finite temperature is also a result of
a varying contribution of the different
spin channels to the 'spin-mixed' electronic states.
In the calculations we can turn the  lattice
vibrations or spin fluctuations  separately on and off.
The main broadening effect in the
spectra results from the spin fluctuations, while lattice vibrations have a minor effect on the spin polarization. However, as it was shown
in the case of soft- and hard-X-ray photoemission \cite{BMM+13} lattice vibrations will become more noticeable at higher photon energies.
 \begin{figure}[h]
  \includegraphics[width=1.0\textwidth,angle=0,clip]{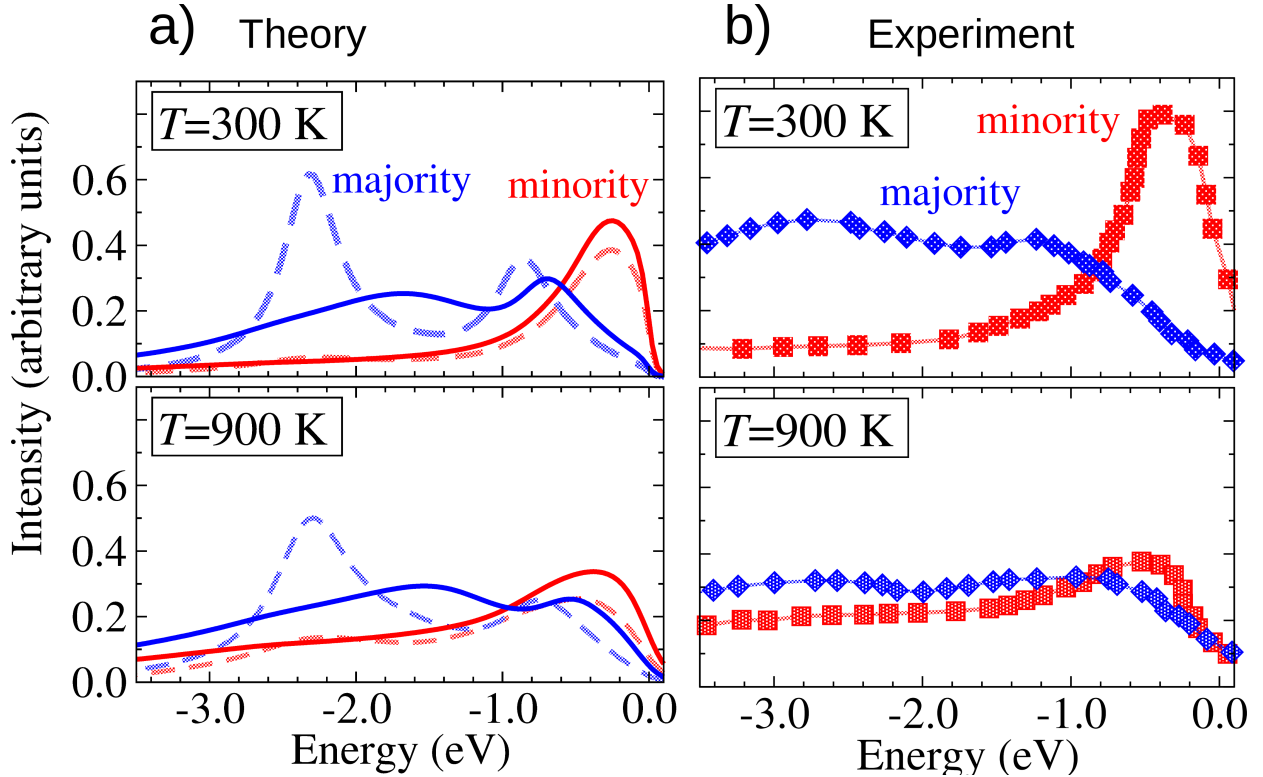}
 \caption{\label{fig:NormalEm_DMFT}
 Comparison between experimental (right panel) and theoretical LSDA+DMFT based calculations (left panel) for temperature dependent
 spin-resolved photoemission as measured for $E_{\rm phot}=60$~eV and
 normal emission.  Dashed lines give calculated spectra obtained by means
 of LSDA (taken from Fig.\ref{fig:NormalEm_LDA}).}
 \end{figure}

It can be seen from Fig.~\ref{fig:NormalEm_LDA}, that the overall agreement between the experimental data and the LSDA based calculations
is quite reasonable. Also the temperature-dependency is well described by the LSDA calculations. However, LSDA based
calculations underestimate the energy-dependent broadening and the position of the E$_{g}$ peak is found at higher binding
energy. One of the most successful approaches to include many-body effects beyond LSDA is the LSDA+DMFT scheme. Various aspects
concerning a self-energy obtained via self-consistent LSDA+DMFT calculations for bcc-Fe have been discussed in detail recently in
the context of ARPES \cite{SFB+09,SBM+12}. To find the best correspondence between the binding-energy positions and energy-dependent
broadening of the theoretical peaks we have used for the averaged on-site Coulomb interaction $U$ the value of $U=1.5$~eV and exchange
$J=0.9$~eV. The chosen value for $U$ lies
 between the estimated value $U \approx 1$~eV
 based on experiment \cite{PhysRevB.45.13272}
 and the value $U\approx 2$~eV derived
from theoretical studies \cite{CG05,CMK+08}. The most pronounced difference between LSDA+DMFT calculations and corresponding
experimental results concerns the majority T$_{2g}$ state which in the LSDA+DMFT calculations is shifted towards the Fermi level.
On the other hand, the energetic position of this peak is better reproduced by plain LSDA calculations as shown in
Fig.~\ref{fig:NormalEm_LDA}. These differences may indicate a strong influence of nonlocal correlations in the case of Fe \cite{SFB+09,SBM+12}.

In the following we address the question to which extent strongly correlated systems can be investigated by means of an implementation
suited to deal with only moderately correlated systems. In general local spin fluctuations and corresponding correlations are formally
included in the LSDA+DMFT calculations if a numerically exact DMFT
impurity solver is used, e.g. by using the continuous time Monte Carlo method.
On the other hand, the spin-polarized T-matrix fluctuation-exchange
solver (SPTF) \cite{PKL05,KL02} used to calculate the spectra presented
in Fig.~\ref{fig:NormalEm_DMFT}, has been implemented to treat the problem of magnetic fluctuations in transition metals, and has been
successfully applied to the ferromagnetic phases of Fe, Co, Ni \cite{KL02,BME+06,GDK+07} and to the anti-ferromagnetic phase of
$\gamma$-Mn \cite{DMB+09}, as well as to half-metallic ferromagnets \cite{KIC+08}. This solver is quite stable, computationally rather
cheap and deals with the complete four-indices interaction matrix. On the other hand, its perturbative character restricts its use to
relatively weakly, or moderately, correlated systems. Not surprisingly, the SPTF performs well when starting from a spin-polarized
solution, since the spin-splitting contains already the main part of the exchange and correlation effects. On the other hand, the
direct application of SPTF to a non-magnetic reference state can create stability problems. This is because one tries to attribute
the strong and essentially mean-field effect of the formation of a local magnetic moment to dynamical fluctuations around the
non-spin-polarized state. Using a non-magnetic reference state causes no problems when one uses the quantum Monte Carlo (MC) method,
which has no formal restrictions on the amplitude of fluctuations, but seems problematic for perturbative approaches. As a way to reduce the
limitations for the latter case we propose a combination of SPTF with the disordered local moment approach \cite{GPS+85,Sta94}. As
already shown for the case of actinides \cite{NWK+03} the inclusion of the fluctuations of randomly oriented local moments can improve
drastically the description of the energetics in the paramagnetic phase. Therefore, as it is demonstrated in Fig.~\ref{fig:NormalEm_DMFT}
one can hope that the combination of spin fluctuations treated within the alloy analogy model presented here in combination with a pertrubative
DMFT solver allows us to extend the range of applicability of SPTF.

 \begin{figure}[h]
  \includegraphics[width=1.0\textwidth,angle=0,clip]{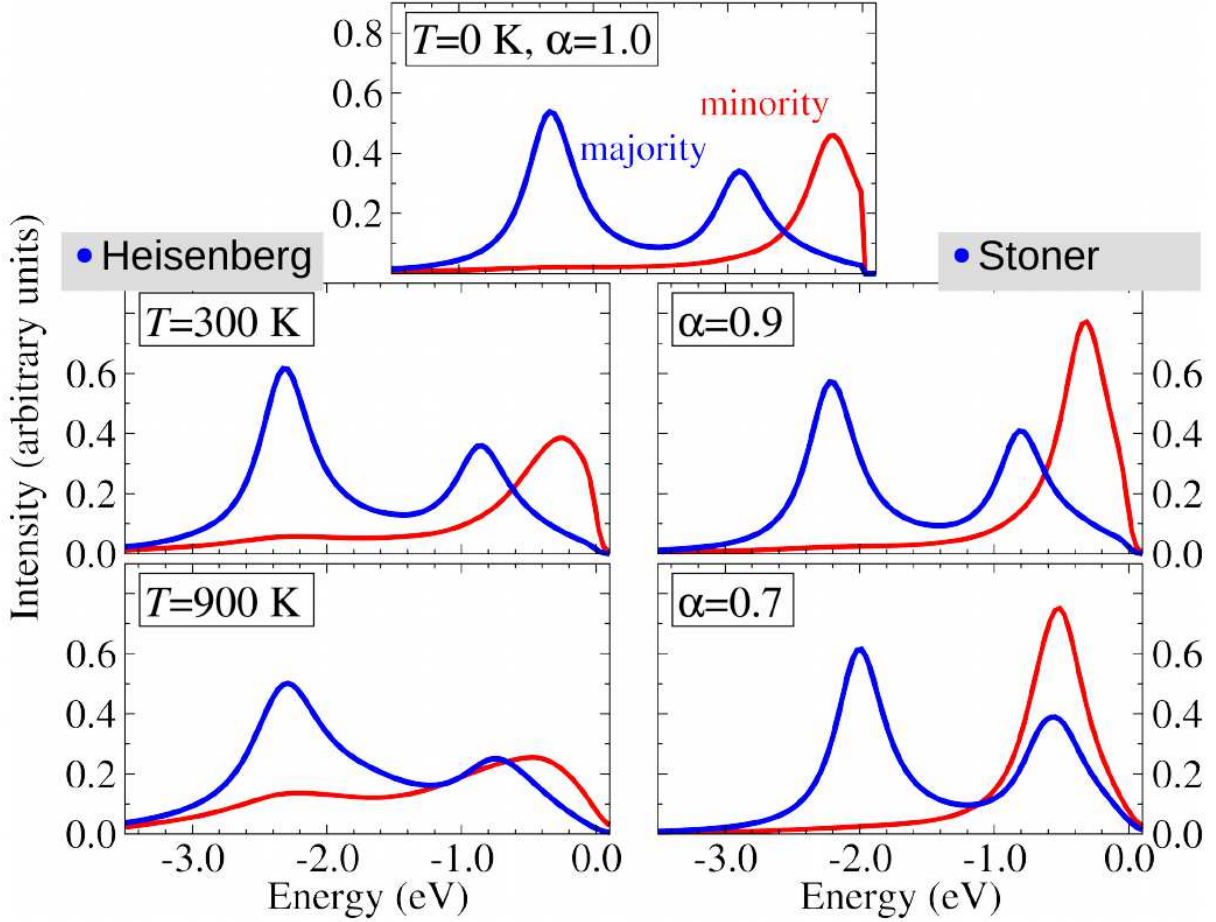}
 \caption{\label{fig:Stoner_Vs_Heisenberg} Calculated spin resolved
   ARPES spectra for $E_{\rm phot}=60$~eV and normal geometry. The results
   in the top panel
  are calculated spectra for $T=0$~K. Bottom left panel: calculated LSDA
  results based on the  alloy analogy model (Heisenberg model).
  Bottom right panel: calculated LSDA results applying a modified exchange splitting (Stoner model).}
 \end{figure}
\begin{figure}[h]
  \includegraphics[width=1.0\textwidth,angle=0,clip]{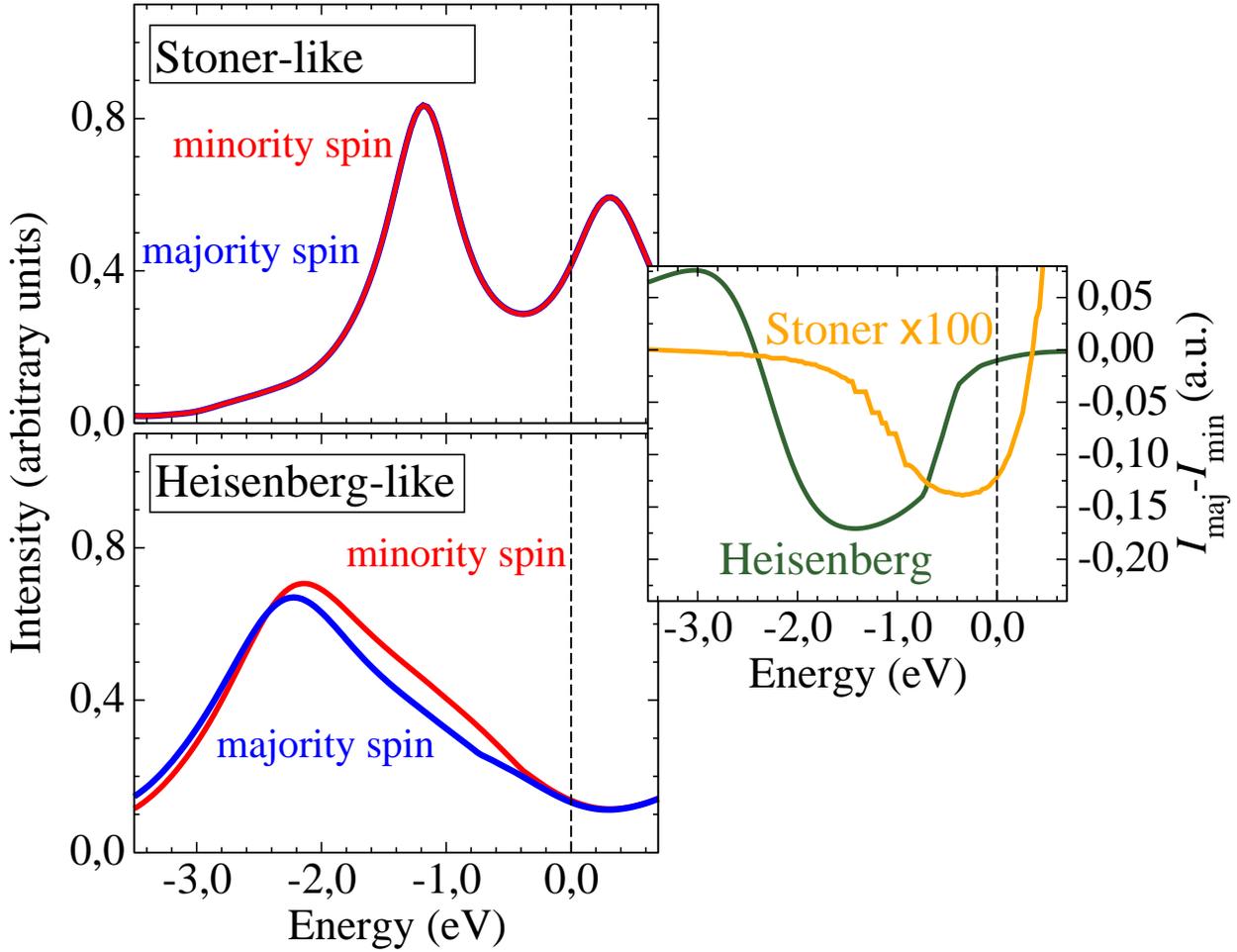}
 \caption{\label{fig:NormalEm_paramag}
   Left panel: Comparison of spin resolved ARPES intensities between Stoner- and Heisenberg-like model
calculated at $T=1100$~K close to ferro- to paramagnetic transition. Right panel: Corresponding spin difference $I_{maj}-I_{min}$.}
 \end{figure}

Within the recent novel ultra fast pump-probe spin-resolved photoemission experiments
on ferromagnetic materials \cite{EPR+17} time-dependent demagnetization is reflected
by a corresponding change in the exchange
splitting. Several mechanisms for this observation  have  been proposed in the literature.
Among others, Eich et al.\ discussed as two possible limiting physical models the
 itinerant electron Stoner-like approach versus the localized electron
Heisenberg spin-fluctuation picture. While the first model allows only for a
homogeneous longitudinal magnetisation in the system, the later one accounts
for transversal spin fluctuations as well. Refering to a common spin quantization
axis in the system, these lead to a band mirroring, i.e.\ to a transfer of
spectral weight of majority- or minority-spin states to mirrored states located close in binding energy but with opposite spin.
Here we point out, that a point of view  similar to  the
band mirroring picture  has been introduced in a more formal way in the past when dealing
with  itinerant ferromagnets at finite tempratures \cite{MJK98,KMP77,Cap74,GPS+85}.
The approach leads to so-called shadow bands and was used among others to discuss
the temperature dependence of  ARPES as well as  magnetoresistance measurements \cite{MJK98}.
Both of these models will lead to
different signatures in the spin-resolved ARPES data and the main question is to what extent are these two models distinguishable by
the use of ab-initio based calculated ARPES spectra. The formalism presented in this manuscript allows to model quantitatively and to
predict in detail all possible differences in the corresponding ARPES spectra. In the left panel of Fig.~\ref{fig:Stoner_Vs_Heisenberg}
we summarize spin-resolved spectra for the Heisenberg model as
calculated by the alloy analogy model for $T=0$, $300$ and $900$~K (results taken
from Fig.~\ref{fig:NormalEm_LDA}). In the right panel, we present calculated spectra for a modified exchange field
$B(\vec r)=\alpha B(\vec r)$, where $\alpha$ is a scaling factor which
has been chosen in such a way that the local magnetic moment
of Fe follows the experimental magnetization curve. We obtain significant differences between the two models. Within the Heisenberg
model the minority-spin channel develops a second peak at higher
binding energy, this way reflecting {\bf the shadow bands and band
  mirroring picture}.
Whereas,
the Stoner model leads to a shift of the minority spin states towards higher binding energies. Finally, as shown in the
Fig.~\ref{fig:NormalEm_paramag}, above  $T_{\rm C}$ the Heisenberg picture still leads to a non-zero spin polarization in the spin-polarized
ARPES spectra due to the photoemission process. On the other hand, the Stoner model leads to zero spin polarization above $T_{\rm C}$ and the
main intensity is found at a binding energy of
about $1$~eV. As a consequence one may state that these explicit spectroscopical calculations
provide an adequate tool to distinguish between the various physical mechanisms involved.

 \section{Conclusions}
 \label{sec:sum}
We have introduced a generalization of the one-step model of photoemission for finite temperatures. The scheme is based on the alloy analogy
model that allows for the inclusion of thermal effects when calculating spin-resolved ARPES spectra. The technical details of the implementation
using the spin-polarized relativistic coherent potential approximation within the one-step model of photoemission have been outlined. This formalism
allows to deal quantitatively with spin-fluctuations as well as with
lattice vibrations on the same footing. In the present
contribution we
have discussed temperature-dependent, spin-resolved ARPES spectra of Fe(001). Our calculated photoemission spectra for Fe(001) were found to match
quantitatively the experimental data.  To
overcome the limitations of local density approximation based
calculations applications of
the LSDA+DMFT scheme have been presented and discussed. The inclusion of electronic correlations described by the pertrubative SPTF-DMFT many
body solver in combination with randomly fluctuating local moments improve the description of the corresponding spectra in the paramagnetic phase.

As it was shown, the alloy analogy model can be used to describe and predict changes of the spin-polarized spectra due to the
ultrafast processes obtained in  pump-probe photoemission. Here we
showed that the Heisenberg like {\bf band mirroring mechanism which leads
to the shadow bands} provide an adequate
model to describe recent experimental findings.

\section{Acknowledgements}
Financial support by DFG (Ebe154/32-1) is thankfully acknowledged.
J.M. would like to thank
CEDAMNF project financed by Ministry of Education, Youth,
and Sports of Czech Rep., Project No. CZ.02.1.01/0.0/0.0/15\_003/0000358.
Authors would like to thank Voicu Popescu for the discussions and his
support when preparing some graphs.


\end{document}